\documentclass[10pt,letterpaper,twocolumn]{article}
\usepackage[T1]{fontenc}
\usepackage[utf8]{inputenc}
\usepackage[margin=0.75in]{geometry}
\usepackage{newtxtext,newtxmath}
\usepackage{graphicx}
\usepackage[hyphens]{url}
\usepackage[authoryear,round]{natbib}
\usepackage{amsmath,bm}
\usepackage{booktabs,multirow}
\usepackage{caption,subcaption}
\usepackage{xcolor}
\usepackage{microtype}
\usepackage[hidelinks]{hyperref}
\let\cite\citep
\AtBeginEnvironment{thebibliography}{\raggedright}
\title{Remix the Timbre: Diffusion-Based Style Transfer Across Polyphonic Stems}
\author{Leduo Chen \qquad Junchuan Zhao\thanks{Corresponding author: Junchuan@u.nus.edu.} \qquad Shengchen Li\\[0.5em]
\small lec015@ucsd.edu, Junchuan@u.nus.edu, Shengchen.Li@xjtlu.edu.cn}
\date{}
\hypersetup{pdftitle={Remix the Timbre: Diffusion-Based Style Transfer Across Polyphonic Stems},pdfauthor={Leduo Chen, Junchuan Zhao, Shengchen Li}}

\begin{document}

\maketitle
\begin{abstract}
Timbre transfer aims to modify the timbral identity of a musical
recording while preserving its original melody and rhythm. While
single-instrument timbre transfer has made substantial progress,
existing approaches to multi-instrument settings commonly rely on
separate-then-transfer pipelines that propagate source-separation
artifacts and process the resulting stems independently. This paper proposes MixtureTT, to the best of our knowledge the first
system for flexible per-stem timbre transfer directly from a
polyphonic mixture. Given a mixture and a separate timbre reference
for each target voice, MixtureTT jointly transfers all stems to the
specified instruments. Its Joint Stem Diffusion Transformer first
models cross-stem harmonic context and then refines each output under
its own content and timbre conditions. All stems are generated within
a single joint diffusion process rather than through separate
per-stem diffusion runs. This design avoids propagating waveform-level
separation errors through a cascade and produces more coherent
multi-stem outputs. Despite operating under a harder input condition, MixtureTT outperforms practical separate-then-transfer pipelines on objective
and subjective evaluations, while remaining competitive with
single-instrument methods given ground-truth isolated stems. Results
on both synthesized and real-audio SATB benchmarks further demonstrate
its generalization. Controlled ablations show that jointly modeling
cross-stem dependencies improves over an equivalent independent-stem
variant, while the structure and timing of this coordination are also
critical for effective mixture-level timbre transfer.
\end{abstract}

\section{Introduction}
\label{sec:introduction}

\begin{figure}[t]
    \centering
    \includegraphics[width=0.9\linewidth]
    {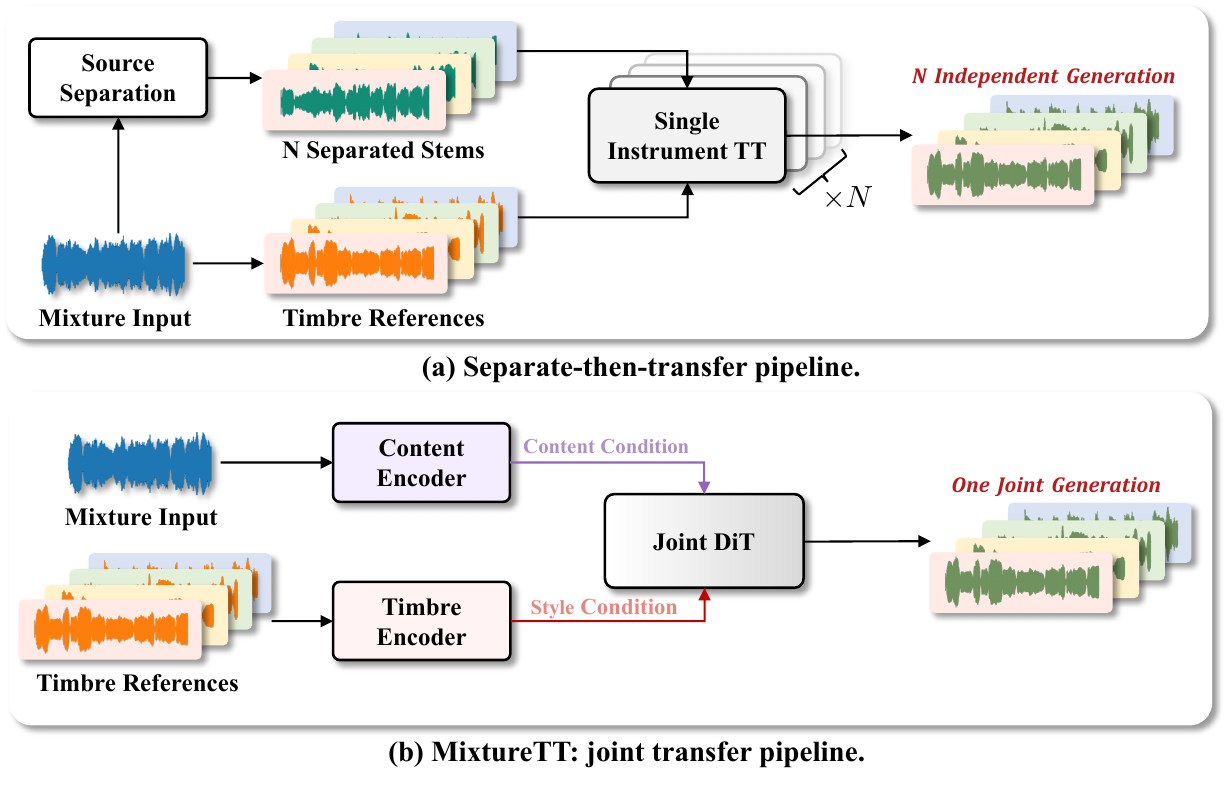}
    \caption{
    \textbf{Top}: the separate-then-transfer pipeline required by
    single-instrument methods.
    \textbf{Bottom}: MixtureTT performs joint per-stem transfer
    directly from the mixture.
    }
    \label{fig:teaser}
\end{figure}

Timbre transfer changes instrumental identity while preserving musical
content such as pitch, rhythm, and articulation. Existing methods
primarily operate on isolated sources, where the content to preserve
is directly observable. The mixture setting is harder: a model must
recover the content of each latent source, apply an independently
specified target timbre, and generate outputs that remain temporally
and harmonically coherent when mixed. Mixture-level timbre transfer
therefore requires both source-specific controllability and explicit
modeling of cross-stem dependencies.

Most timbre-transfer systems assume isolated audio
~\cite{engel2017neural,mor2018universal,huang2018timbretron,
cifka2021self,comanducci2023timbre,demerle2024combining,
mancusi2025latent}. Methods operating directly on mixtures typically
transform the ensemble as a single timbral unit
~\cite{alinoori2022music,baoueb2024wavetransfer}, or require external
queries to identify selected sources~\cite{luo2024dismix}. A practical
alternative is \emph{separate-then-transfer}, which estimates each
waveform and applies a single-instrument model independently. This
pipeline propagates separation artifacts and treats the output stems
as unrelated generation problems, providing no explicit mechanism for
preserving their shared temporal and harmonic structure. Its contrast
with MixtureTT is illustrated in Figure~\ref{fig:teaser}.

Joint generation offers a natural alternative, but simply allowing the
stems to interact is insufficient. Independent processing ignores
ensemble dependencies, whereas unrestricted interaction between
evolving noisy stems may introduce cross-talk and blur source
identity. The central questions are therefore not only
\emph{whether} the stems should be modeled jointly, but also
\emph{through what representation} and \emph{when} their interaction
should occur. We hypothesize that the stems should first establish a
shared interpretation of the musical event through source-aware
content, and only then specialize under their individual content and
target-timbre conditions.

We propose \textbf{MixtureTT}, a latent diffusion system for
independently controlled per-stem timbre transfer directly from a
polyphonic mixture. In the fixed-source setting considered here,
MixtureTT jointly denoises all target stems within a shared diffusion
process. A pretrained HT Demucs encoder~\cite{rouard2023hybrid}
factorizes the mixture into source-specific content representations
without decoding intermediate source waveforms. These representations
condition their corresponding outputs and form an
\textbf{Ensemble-Content Bus (ECB)}, which mediates early cross-stem
coordination before stem-private refinement. This
\emph{factorize--coordinate--specialize} design models ensemble
dependencies while preserving source-specific content and independent
target-timbre control.

Experiments on CocoChorales show that MixtureTT outperforms practical
separate-then-transfer pipelines and remains competitive with
single-instrument methods given ground-truth stems. It also transfers
without fine-tuning to ChoraleBricks mixtures constructed from real
multitrack recordings. Controlled ablations show that joint denoising
improves over matched independent-stem generation, while unrestricted
interaction and entangled mixture conditioning are less effective
than early, content-mediated coordination. Listening tests further
support improvements in perceptual quality and inter-stem coherence.

Our contributions are:
\begin{itemize}
    \item We introduce MixtureTT, to our knowledge the first system for
    independently controlled per-stem timbre transfer directly from a
    polyphonic mixture, without materializing intermediate
    waveform-level source estimates.

    \item We develop a Joint Stem Diffusion framework that explicitly
    models cross-stem dependencies while preserving source-specific
    content and independent target-timbre control. Its
    coordinate-then-specialize architecture mediates early interaction
    through source-aware ensemble content.

    \item Through separated- and ground-truth-stem comparisons,
    controlled interaction ablations, stage-order analysis, and
    subjective evaluation, we show that joint denoising is functionally
    important and that effective coordination depends on both its
    representation and timing.
\end{itemize}

\section{Related Work}

\subsection{Single-Instrument Timbre Transfer}

Single-instrument timbre transfer has progressed from waveform and
domain translation~\cite{engel2017neural,mor2018universal,
huang2018timbretron} and differentiable synthesis
~\cite{engel2020ddsp,hayes2021neural,caspe2022ddx7,wu2022mididdsp} to
factorized, adversarial, and discrete latent
representations~\cite{hung2018learning,wu2023transplayer,
cifka2021self,bitton2020vector,bonnici2022timbre,esling2018generative,
luo2019learning}.
Early translation
methods demonstrated cross-instrument conversion but typically rely
on fixed source--target domains, while DDSP provides explicit pitch
and loudness control under a primarily monophonic harmonic-source
assumption. Latent-variable methods improve many-to-many or
reference-conditioned transfer, and recent diffusion approaches
further enhance generation quality through explicit structural and
timbral conditioning~\cite{demerle2024combining,mancusi2025latent}.

These methods nevertheless treat the input as a single timbral
entity, even when it contains a polyphonic performance of one
instrument. MixtureTT addresses the complementary problem of applying
independent reference timbres to multiple sources within the same
mixture while generating them jointly.

\subsection{Mixture-Level Timbre Manipulation}

Existing mixture-level methods primarily perform global
re-instrumentation. Music-STAR~\cite{alinoori2022music},
DiffTransfer~\cite{comanducci2023timbre}, and
WaveTransfer~\cite{baoueb2024wavetransfer} transform a complete
mixture or predefined ensemble configuration into another. Although
they avoid isolated source inputs, their target mappings are generally
fixed during training and do not provide independent reference-based
control over the constituent voices.

DisMix~\cite{luo2024dismix} enables finer source manipulation through
query-conditioned pitch and timbre representations, but requires an
external recording to identify each source. Multi-source diffusion
models~\cite{mariani2023multi,postolache2024generalized,
karchkhadze2025simultaneous,parker2024stemgen} jointly model coherent stems,
yet focus on generation, separation, or source imputation rather than
content-preserving re-instrumentation.

Existing approaches therefore offer either global mixture
transformation or source-level manipulation with additional queries.
MixtureTT instead provides independent audio-reference control for
all sources directly from a mixture, without parallel cross-timbre
renditions, source-identification queries, or intermediate
waveform-level estimates.

\subsection{Diffusion Models and Training-Free Music Editing}

Diffusion and latent-diffusion models
~\cite{ho2020denoising,song2020score,rombach2022high} underpin modern
audio generation~\cite{liu2023audioldm,evans2025stable}. MixtureTT
uses EDM preconditioning~\cite{karras2022elucidating,evans2025stable}
and applies classifier-free guidance~\cite{ho2022classifier}
separately to content and timbre.

Training-free editors instead manipulate pretrained diffusion models
through text or latent editing~\cite{zhang2024musicmagus}, diffusion
inversion~\cite{manor2024zero}, score
distillation~\cite{niu2026steermusic}, or reference-feature
injection~\cite{wang2024stylus}. Stylus~\cite{wang2024stylus}
supports reference-based style transfer but edits the mixture
globally, while Polyphonia~\cite{li2026polyphonia} modifies one
selected source while preserving its accompaniment. These methods
avoid task-specific training, but do not jointly re-instrument all
sources under independent references. We therefore include Stylus as
a training-free mixture-level baseline.

\begin{figure*}[t]
    \centering
    \includegraphics[width=0.93\linewidth]{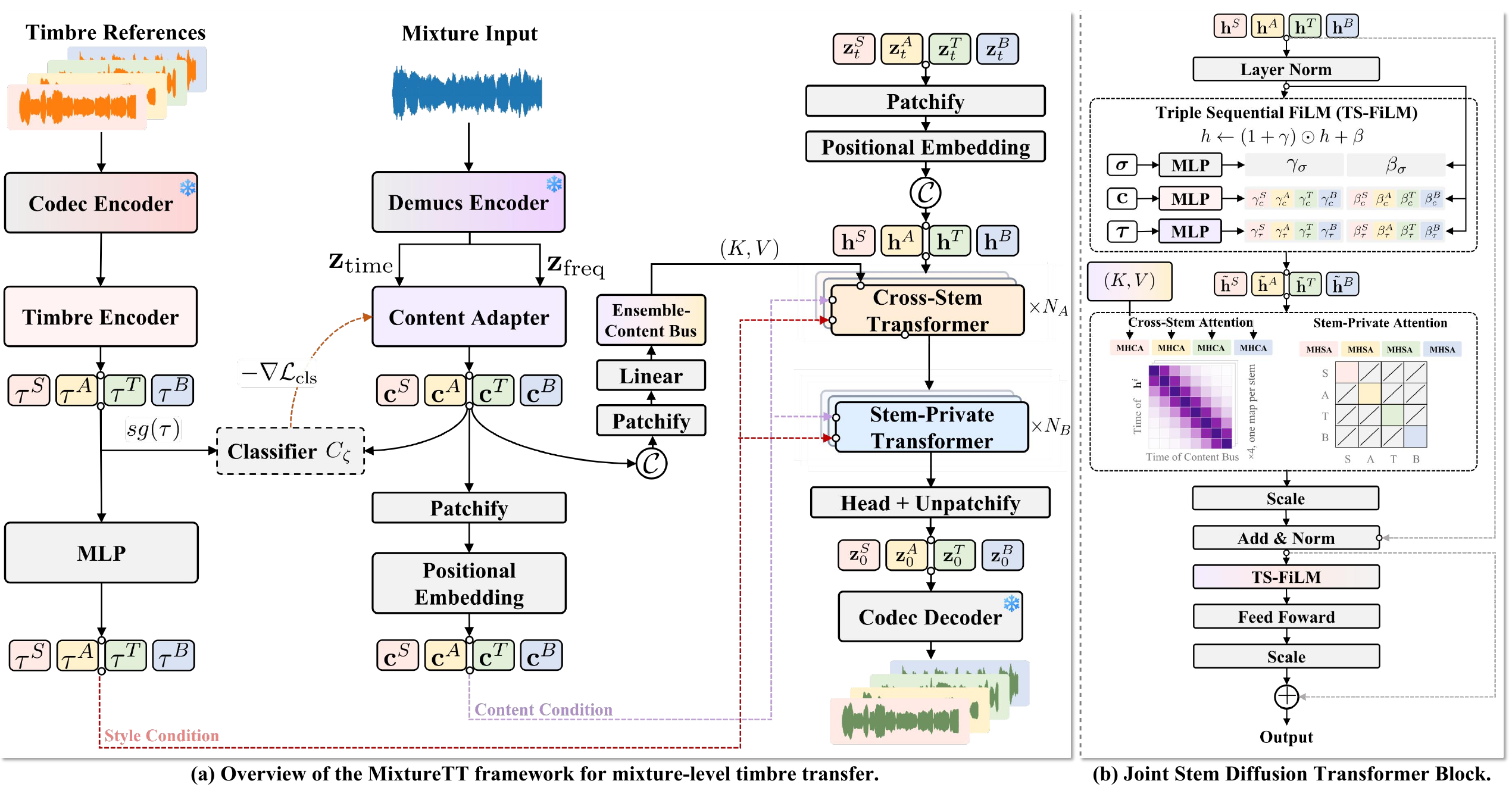}
    \caption{
Given a mixture and one target-timbre reference per source, MixtureTT follows a \emph{factorize--coordinate--specialize} pipeline. A frozen HT Demucs encoder and a trainable dual-branch adapter first extract source-specific content without decoding intermediate waveforms. The Joint Stem DiT then jointly generates all target stems using a \emph{coordinate-then-specialize} design: two early blocks mediate cross-stem dependencies through an Ensemble-Content Bus (ECB), after which stem-private blocks refine each output under its own content and target-timbre conditions. A frozen codec decoder reconstructs the generated waveforms.
    }
    \label{fig:overview}
\end{figure*}

\section{Method} 
\label{sec:method} 
MixtureTT receives one polyphonic mixture and one target-timbre reference for each of $N$ source slots, and jointly generates all output stems in latent space. It follows a \emph{factorize--coordinate--specialize} design: a source-aware pathway first factorizes the mixture into stem-specific content representations without reconstructing intermediate waveforms; the Joint Stem DiT then coordinates the stems through an Ensemble-Content Bus (ECB) before refining them with stem-private blocks. This structure preserves shared temporal and harmonic organization while maintaining source-specific content and independent target-timbre control. Figure~\ref{fig:overview} summarizes the architecture.

\subsection{Audio Codec}
\label{ssec:codec}

Neural codecs and autoencoders provide compact latent spaces for audio
compression and generation~\cite{zeghidour2021soundstream,defossez2022high,
kumar2023high}. Following CTD~\cite{demerle2024combining}, a RAVE-style
convolutional codec~\cite{caillon2021rave} is adopted with the adversarial
discriminator of~\cite{kumar2023high}, consistent with common practice in
latent-domain generation and conversion~\cite{liu2023audioldm,yang2023uniaudio,
wang2025discl,zhao25d_interspeech,zhao2026comelsinger}. The codec encodes each
waveform $\mathbf{x}$ into $\mathbf{z}\in\mathbb{R}^{L\times D_z}$ and is
pretrained and frozen, so diffusion operates in the latent space and the decoder
reconstructs waveforms at inference.


\subsection{Mixture-Conditioned Content Factorization}
\label{ssec:content}
MixtureTT factorizes source content directly in representation space
rather than estimating intermediate source waveforms. We use the
frozen encoder of a pretrained HT Demucs separator
~\cite{rouard2023hybrid} as a source-aware acoustic prior, while
discarding its separation decoder.

Following hybrid waveform--spectrogram modeling in music source
separation~\cite{defossez2021hybrid,rouard2023hybrid}, a trainable Dual-Branch Content Adapter combines the encoder's
frequency- and time-domain bottleneck features. The frequency branch
is compressed along the spectral axis and temporally aligned with the
time-domain branch, after which cross-attention~\cite{vaswani2017attention} injects spectral
evidence into the temporal representation. The fused feature is then
combined with the original temporal stream and decoded by $N$
source-specific heads, producing content sequences
$\{\mathbf{c}^{(i)}\}_{i=1}^{N}$ from a single mixture encoding.

Because independently decoded heads may assign overlapping mixture
evidence inconsistently, a lightweight source-axis Transformer jointly
refines the $N$ representations at each temporal position. This lets
the heads resolve competing source explanations before generation.
The operation is applied once during content factorization and is
distinct from the Joint Stem DiT, which coordinates the evolving noisy
stems throughout denoising.

MixtureTT therefore avoids waveform-level separation as an
intermediate inference step, but remains dependent on a source-aware
content backbone whose source taxonomy matches the target task.

\subsection{Reference Timbre Encoding and Disentanglement}
\label{ssec:timbre}

\paragraph{Reference timbre encoding.}
Each target-reference waveform is first mapped to the frozen codec
space and then processed by a trainable convolutional timbre encoder
with global pooling, yielding a time-invariant embedding
$\boldsymbol{\tau}^{(i)}$. During training, the reference is sampled
from a different temporal window of the same instrument track, so it
shares instrument identity with the target but not its local melody or
rhythm. This encourages the embedding to represent global timbre
rather than source content. At inference, the reference can be any
suitable recording of the desired target instrument.
\paragraph{Adversarial content--timbre disentanglement.}
A low-dimensional content bottleneck does not guarantee that
instrument identity has been removed from
$\mathbf{c}^{(i)}$. We therefore introduce an auxiliary predictor
$C_{\zeta}$ that attempts to recover the timbre embedding from the
content sequence:
\begin{equation}
\begin{aligned}
    \mathcal{L}_{\mathrm{cls}}
    =
    \frac{1}{N}
    \sum_{i=1}^{N}
    \Bigl\|
        C_{\zeta}
        \bigl(
            \mathbf{c}^{(i)}
        \bigr)
        -
        \operatorname{sg}
        \bigl(
            \boldsymbol{\tau}^{(i)}
        \bigr)
    \Bigr\|_2^2 ,
\end{aligned}
\label{eq:cls}
\end{equation}
where $\operatorname{sg}(\cdot)$ denotes stop-gradient.

The two sides of this adversarial objective are optimized in opposite
directions. The predictor parameters $\zeta$ minimize
$\mathcal{L}_{\mathrm{cls}}$ and therefore learn to detect residual
timbre information in the content representation. In contrast, the
content adapter receives the reversed objective
$-\lambda_{\mathrm{cls}}\mathcal{L}_{\mathrm{cls}}$, making timbre
prediction increasingly difficult. This gradient-reversal-style
training explicitly purifies the content space instead of relying on
bottleneck dimensionality alone.

\paragraph{Timbre anti-collapse regularization.}
Our training protocol assigns distinct target instruments to the
$N$ stems. To prevent the timbre encoder from collapsing, we penalize
high similarity between their reference embeddings:
\begin{equation}
\mathcal{L}_{\mathrm{div}}^{\mathrm{cross}}
=
\tfrac{1}{|\mathcal{P}|}
\sum_{(i,j)\in\mathcal{P}}
\cos^2\!\left(
\boldsymbol{\tau}^{(i)},\boldsymbol{\tau}^{(j)}
\right),
\label{eq:div_cross}
\end{equation}
where $\mathcal{P}$ contains pairs assigned to different target
instruments. We additionally apply variance regularization, following
explicit anti-collapse objectives~\cite{bardes2021vicreg}:
\begin{equation}
\mathcal{L}_{\mathrm{div}}^{\mathrm{var}}
=
\tfrac{1}{N}
\sum_{i=1}^{N}
\max\!\left(
0,\delta-\mathrm{std}_{\mathcal{B}}
(\boldsymbol{\tau}^{(i)})
\right),
\label{eq:div_var}
\end{equation}
where $\mathrm{std}_{\mathcal{B}}$ denotes batch-wise standard
deviation and $\delta$ is a target threshold. The total regularizer is
$\mathcal{L}_{\mathrm{div}}
=
\mathcal{L}_{\mathrm{div}}^{\mathrm{cross}}
+
\lambda_{\mathrm{var}}
\mathcal{L}_{\mathrm{div}}^{\mathrm{var}}$.
Both penalties are used only during training.
\subsection{Coordinate-then-Specialize Joint Stem DiT}
\label{ssec:dit}

The central component of MixtureTT is a Joint Stem Diffusion
Transformer that updates all $N$ output stems within the same
denoiser evaluation. To support ensemble coordination without
blurring source identity, the transformer separates cross-stem
interaction from source-specific generation: two early coordination
blocks are followed by eight stem-private refinement blocks.

\paragraph{Conditioning and tokenization.}
At noise level $\sigma$, each noisy source latent is patchified along
time and projected into a token sequence
$\mathbf{h}^{(i)}$. Its source-content sequence is projected at the
same temporal resolution, while the global target-timbre embedding is
broadcast across time. Because diffusion time, content, and timbre
play different conditioning roles, each is injected through a
separate adaptive modulation pathway. This prevents the temporally
dense content condition from dominating the global timbre signal.

\paragraph{Ensemble-content coordination.}
The source-content sequences represent complementary components of the
same musical event. We concatenate them along the channel dimension
and project the result into the Ensemble-Content Bus (ECB):
\begin{equation}
\mathbf{e}_m
=
\operatorname{Proj}_m
\left(
\operatorname{Concat}_{\mathrm{ch}}
\left[
\mathbf{c}^{(1)},\ldots,\mathbf{c}^{(N)}
\right]
\right),
\label{eq:content_bus}
\end{equation}
where each token of $\mathbf{e}_m$ at time $t$ summarizes the
source-specific content of the complete ensemble at that time. The ECB
is the only explicit shared coordination interface in the diffusion
backbone: it contains neither target-timbre conditions nor raw mixture
codec latents, and remains fixed with respect to the evolving noisy
stem states.

In the first two blocks, each stem queries the ECB:
\begin{equation}
\mathbf{a}^{(i)}
=
\operatorname{MHA}
\left(
\Phi(\mathbf{h}^{(i)}),
\mathbf{e}_m,
\mathbf{e}_m
\right),
\label{eq:cross_stem}
\end{equation}
where $\Phi$ denotes the separate modulation by diffusion time,
source content, and target timbre. The stems share the ECB as keys and
values while retaining distinct conditioned queries, allowing each
stem to retrieve synchronized ensemble context without attending
directly to another noisy stem. Performing this coordination first
establishes shared temporal and harmonic structure before
stem-private specialization.

\paragraph{Stem-private specialization.}
After coordination, the remaining blocks use block-diagonal
self-attention. Each stem can attend across its own temporal sequence,
but no further cross-stem information flow is permitted. Most of the
transformer capacity is thus devoted to realizing each source under
its own content and target-timbre conditions after ensemble context
has already been established.

A final modulated output head maps each sequence back to codec-latent
patches, which are unpatchified to obtain the denoised outputs
$\{\widehat{\mathbf{z}}_{0}^{(i)}\}_{i=1}^{N}$. During training, the
stem states and their corresponding content and timbre conditions are
jointly permuted and restored after prediction, preventing the model
from using a fixed slot position as a shortcut for source identity.


\subsection{Joint Objective and Progressive Optimization}
\label{ssec:loss}

Following EDM~\cite{karras2022elucidating}, each clean source latent
is perturbed as
\begin{equation}
    \mathbf{z}_{t}^{(i)}
    =
    \mathbf{z}_{0}^{(i)}
    +
    \sigma
    \boldsymbol{\epsilon}^{(i)},
    \qquad
    \boldsymbol{\epsilon}^{(i)}
    \sim
    \mathcal{N}
    \left(
        \mathbf{0},
        \mathbf{I}
    \right).
    \label{eq:forward_noise}
\end{equation}
All stems from the same example share the noise level $\sigma$ and
are predicted jointly within one denoiser evaluation. The diffusion
objective is
\begin{equation}
\begin{aligned}
    \mathcal{L}_{\mathrm{diff}}
    =
    \mathbb{E}
    \left[
        \frac{w(\sigma)}{N}
        \sum_{i=1}^{N}
        \left\|
            \widehat{\mathbf{z}}_{0}^{(i)}
            -
            \mathbf{z}_{0}^{(i)}
        \right\|_2^2
    \right],
\end{aligned}
\label{eq:diff_loss}
\end{equation}
where $w(\sigma)$ is the standard EDM loss weighting.

The content adapter, timbre encoder, and joint denoiser are optimized
with
\begin{equation}
    \mathcal{L}
    =
    \mathcal{L}_{\mathrm{diff}}
    -
    \lambda_{\mathrm{cls}}
    \mathcal{L}_{\mathrm{cls}}
    +
    \lambda_{\mathrm{div}}
    \mathcal{L}_{\mathrm{div}},
    \label{eq:total_loss}
\end{equation}
while the auxiliary predictor $C_{\zeta}$ is optimized separately to
minimize $\mathcal{L}_{\mathrm{cls}}$.

During joint training, content and timbre conditions are independently
dropped, with content dropped more frequently to prevent the model
from ignoring the reference embedding. The ECB is also occasionally
dropped to prevent excessive dependence on shared ensemble context. Exact architecture, training schedules, condition-dropout probabilities, and optimization details are provided in the supplementary material.

\begin{table*}[t]
\centering
\caption{
Objective evaluation on CocoChorales and ChoraleBricks under
within-instrument reconstruction (R) and cross-timbre transfer (T) settings.
For the single-instrument baselines, \emph{Sep.} denotes the practical
separate-then-transfer pipeline on separated stems, while \emph{GT}
provides ground-truth isolated stems. Stylus and MixtureTT operate directly on
the polyphonic mixture (\emph{Mix.}).
}
\label{tab:main_results}
\normalsize
\setlength{\tabcolsep}{0pt}
\renewcommand{\arraystretch}{1.0}
\setlength{\aboverulesep}{1pt}
\setlength{\belowrulesep}{1pt}
\setlength{\abovetopsep}{0pt}
\setlength{\belowbottomsep}{0pt}

\begin{tabular*}{\textwidth}{@{\extracolsep{\fill}}llcccccccccccc@{}}
\toprule

\multirow{3}{*}{\textbf{Method}} &
\multirow{3}{*}{\textbf{Input}} &
\multicolumn{8}{c}{\textbf{Per-Stem}} &
\multicolumn{4}{c}{\textbf{Mixture}} \\

\cmidrule(lr){3-10}\cmidrule(l){11-14}

& &
\multicolumn{2}{c}{FAD$\downarrow$} &
\multicolumn{2}{c}{JD$\downarrow$} &
\multicolumn{2}{c}{MFCC-cos$\downarrow$} &
\multicolumn{2}{c}{Conf$\uparrow$} &
\multicolumn{2}{c}{FAD$_m\!\downarrow$} &
\multicolumn{2}{c}{CCS$\uparrow$} \\

\cmidrule(lr){3-4}\cmidrule(lr){5-6}\cmidrule(lr){7-8}
\cmidrule(lr){9-10}\cmidrule(lr){11-12}\cmidrule(l){13-14}

& & R & T & R & T & R & T & R & T & R & T & R & T \\

\midrule

\multicolumn{14}{@{}l}{\textbf{\textit{CocoChorales Dataset~\cite{wu2022chamber}}}} \\
\midrule
\multirow{2}{*}{SS-VAE~\cite{cifka2021self}}
& Sep.
& 0.173 & 0.187 & 0.222 & 0.277 & 0.011 & 0.040
& 0.981 & \underline{0.978} & 0.112 & 0.113 & 0.903 & 0.894 \\
& GT
& 0.201 & 0.204 & 0.204 & 0.302 & 0.028 & 0.047
& \underline{0.984} & 0.830 & 0.135 & 0.132 & \underline{0.961} & 0.896 \\

\multirow{2}{*}{CTD~\cite{demerle2024combining}}
& Sep.
& 0.073 & 0.087 & 0.190 & 0.205 & 0.017 & \underline{0.021}
& 0.825 & 0.658 & 0.060 & 0.064 & 0.938 & 0.930 \\
& GT
& \underline{0.053} & \underline{0.062} & \underline{0.188} & \underline{0.204}
& \underline{0.015} & \textbf{0.021}
& 0.809 & 0.621 & \underline{0.055} & \underline{0.058} & 0.951 & \underline{0.931} \\

Stylus~\cite{wang2024stylus}
& Mix.
& \multicolumn{8}{c}{\textit{n/a}}
& 0.242 & 0.211 & 0.884 & 0.884 \\

\textbf{MixtureTT (Ours)}
& Mix.
& \textbf{0.010} & \textbf{0.018} & \textbf{0.157} & \textbf{0.199}
& \textbf{0.005} & 0.033 & \textbf{0.989} & \textbf{0.979}
& \textbf{0.007} & \textbf{0.014} & \textbf{0.985} & \textbf{0.937} \\

\midrule

\multicolumn{14}{@{}l}{\textbf{\textit{ChoraleBricks Dataset~\cite{balke2025choralebricks}}}} \\
\midrule
\multirow{2}{*}{SS-VAE~\cite{cifka2021self}}
& Sep.
& 0.544 & 0.566 & 0.665 & 0.668 & 0.047 & 0.049
& \underline{0.808} & \underline{0.801} & 0.282 & 0.287 & 0.791 & 0.797 \\
& GT
& 0.555 & 0.542 & \underline{0.495} & \underline{0.510}
& \underline{0.043} & \underline{0.052}
& \textbf{0.821} & 0.778 & 0.277 & 0.280 & 0.835 & 0.837 \\

\multirow{2}{*}{CTD~\cite{demerle2024combining}}
& Sep.
& 0.529 & 0.557 & 0.702 & 0.726 & 0.058 & 0.067
& 0.689 & 0.666 & 0.234 & 0.221 & 0.854 & 0.855 \\
& GT
& \underline{0.451} & \underline{0.476} & 0.606 & 0.606 & 0.053 & 0.062
& 0.713 & 0.659 & \underline{0.228} & \underline{0.212}
& \underline{0.869} & \underline{0.865} \\

Stylus~\cite{wang2024stylus}
& Mix.
& \multicolumn{8}{c}{\textit{n/a}}
& 0.735 & 0.723 & 0.850 & 0.849 \\


\textbf{MixtureTT (Ours)}
& Mix.
& \textbf{0.414} & \textbf{0.412} & \textbf{0.464} & \textbf{0.460}
& \textbf{0.038} & \textbf{0.039} & 0.779 & \textbf{0.807}
& \textbf{0.217} & \textbf{0.208} & \textbf{0.874} & \textbf{0.867} \\

\bottomrule
\end{tabular*}
\end{table*}

\section{Experiments}
\label{sec:experiments}

\paragraph{Datasets and comparison protocols.}
Unless otherwise stated, all learned models are trained on the 16-kHz
\emph{tiny} partition of CocoChorales~\cite{wu2022chamber}, containing
24k/8k/8k train/validation/test SATB examples with isolated stems and
mixtures. We additionally evaluate zero-shot transfer without
fine-tuning on ChoraleBricks~\cite{balke2025choralebricks}, using
mixtures constructed from real multitrack recordings. We compare
MixtureTT with SS-VAE~\cite{cifka2021self} and
Control-Transfer-Diffusion (CTD)~\cite{demerle2024combining}, both
retrained on CocoChorales using their official implementations. Since
these methods accept only isolated sources, they are evaluated using
either HT Demucs-separated stems (\emph{Sep.}), representing a
practical separate-then-transfer cascade, or ground-truth isolated
stems (\emph{GT}), providing a privileged input condition.
Stylus~\cite{wang2024stylus} follows its original training-free
mixture-to-mixture protocol and is evaluated only at the mixture level
because it does not produce isolated stems. To study dependence on
paired supervision, we also progressively replace paired
CocoChorales examples $\mathcal{D}_s$ with external mixtures
$\mathcal{D}_u$ pseudo-labeled by HT Demucs, from fully paired to fully
pseudo-labeled training while keeping the total data and optimization
budgets fixed. Dataset construction, training details, and
pseudo-labeling details are provided in the supplementary material.

\paragraph{Controlled ablations.}
All architectural variants retain the same content and timbre
pathways, DiT capacity, and training budget. \emph{Independent stems}
removes inter-stem interaction, testing whether joint denoising is
beneficial. \emph{Full self-attention} instead allows direct
interaction among all noisy-stem tokens, while \emph{mixture
injection} replaces the Ensemble-Content Bus with the frozen codec
latent of the input mixture. These variants compare independent
generation, unrestricted interaction, and coordination through an
entangled mixture representation. To study interaction order, the
same two coordination blocks are placed before, between, or after the
eight stem-private blocks. Finally, removing
$\mathcal{L}_{\mathrm{cls}}$ or $\mathcal{L}_{\mathrm{div}}$ evaluates
content--timbre leakage and target-embedding collapse, respectively.

\paragraph{Evaluation.}
We evaluate within-instrument reconstruction (R) and cross-timbre
transfer (T). Audio quality is measured by per-stem Fr\'echet Audio
Distance (FAD) and mixture-level FAD (FAD$_m$); content preservation
by pitch-contour Jaccard Distance (JD) and mixture-level Chroma Cosine
Similarity (CCS); and timbre control by MFCC cosine distance
(MFCC-cos) and target-instrument classifier confidence (Conf.). An
anonymous study with 35 participants further evaluates transfer
success, content preservation, sound quality, and inter-stem
coherence using five-point Mean Opinion Score (MOS) ratings. Full
evaluator and listening-test protocols are provided in the
supplementary material.

\section{Results}
\label{sec:results}

\subsection{Main Results}
\label{ssec:main_results}

\begin{figure}[t]
\centering
\setlength{\tabcolsep}{1pt}
\renewcommand{\arraystretch}{0.6}
\begin{tabular}{c@{\hskip 2pt}ccccc}
 & \scriptsize Input mix & \scriptsize GT flute & \scriptsize GT oboe & \scriptsize GT clarinet & \scriptsize GT bassoon \\
\rotatebox{90}{\scriptsize\textbf{Source}} &
\includegraphics[width=0.16\linewidth]{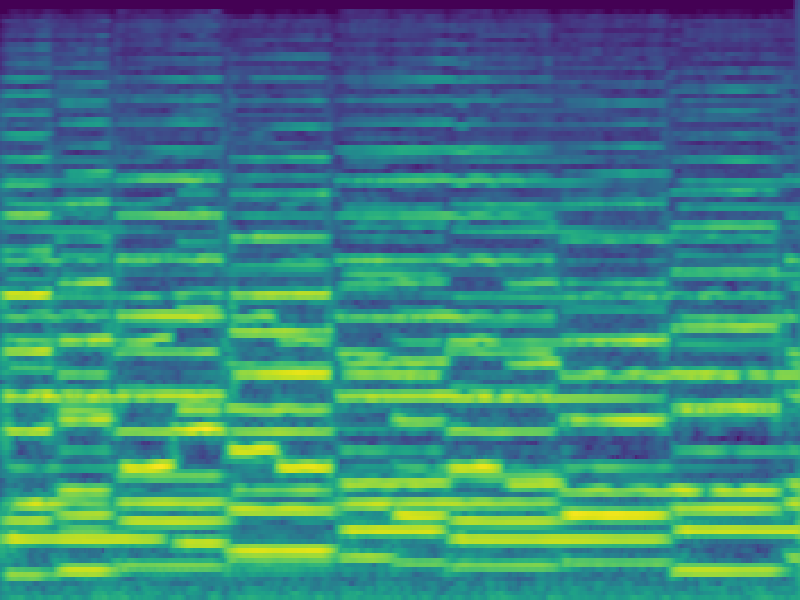} &
\includegraphics[width=0.16\linewidth]{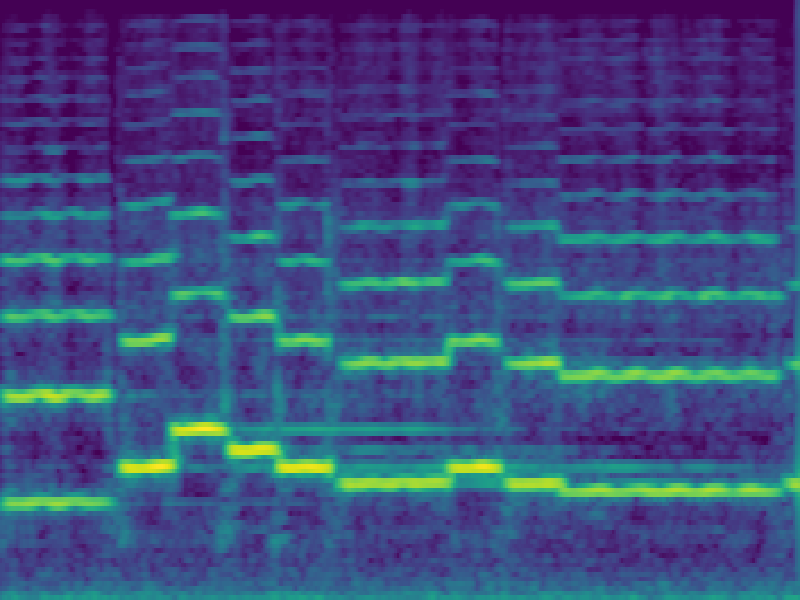} &
\includegraphics[width=0.16\linewidth]{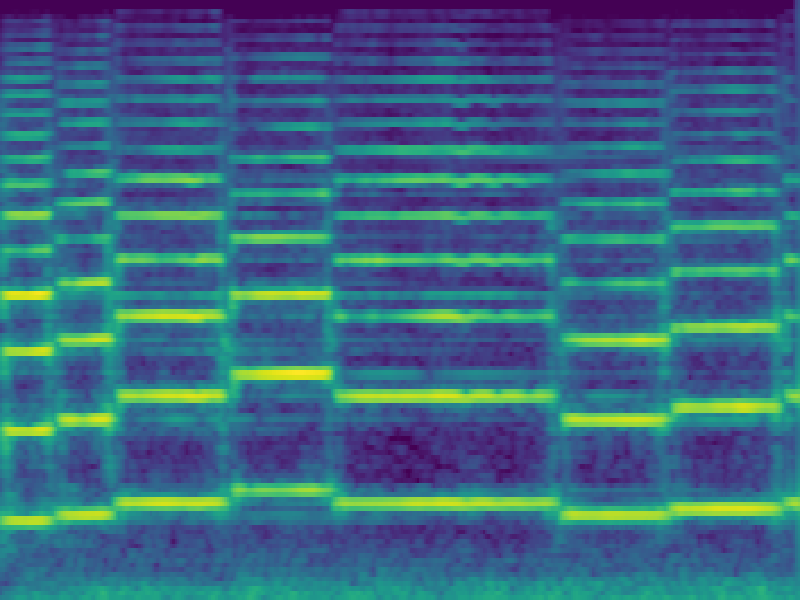} &
\includegraphics[width=0.16\linewidth]{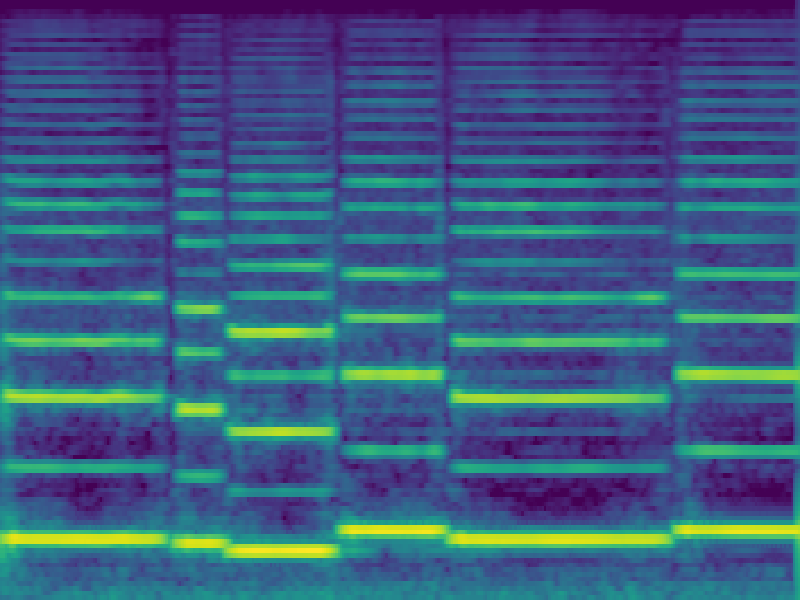} &
\includegraphics[width=0.16\linewidth]{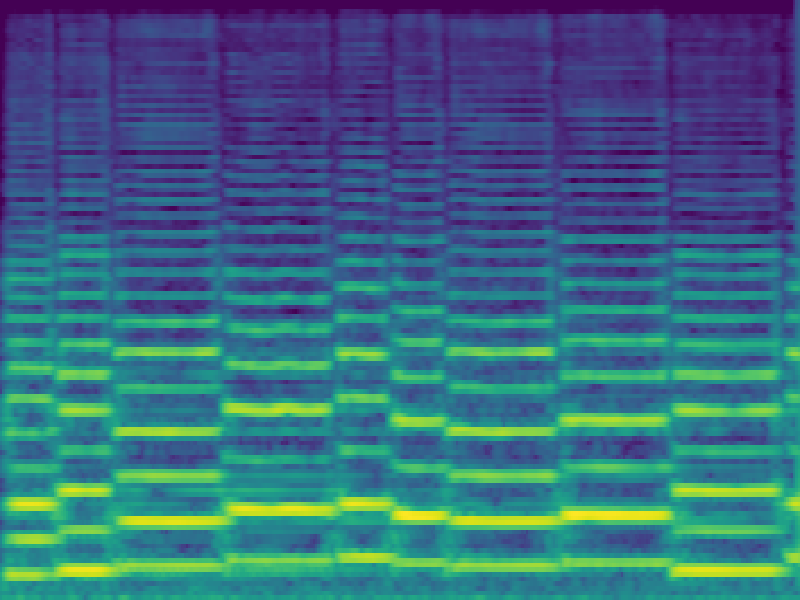} \\[2pt]
 & \scriptsize GT mix & \scriptsize Ref trumpet & \scriptsize Ref horn & \scriptsize Ref trombone & \scriptsize Ref tuba \\
\rotatebox{90}{\scriptsize\textbf{Ref.}} &
\includegraphics[width=0.16\linewidth]{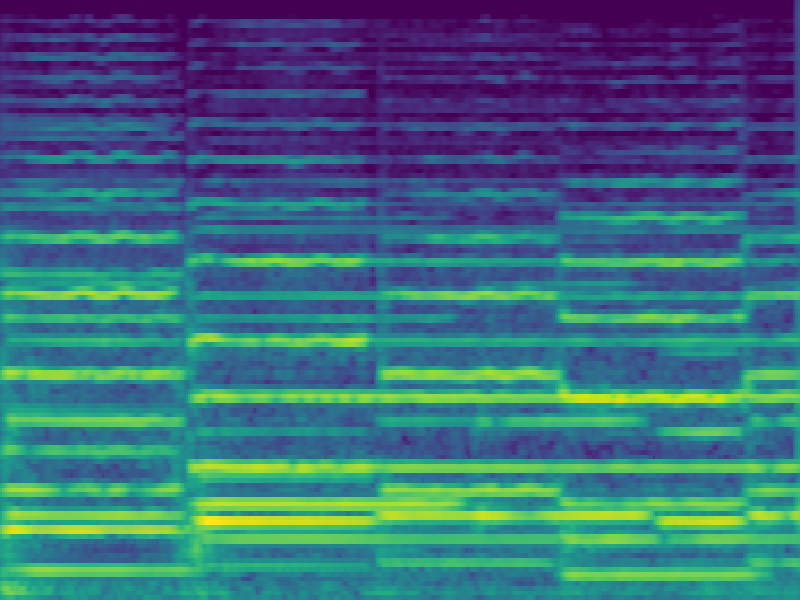} &
\includegraphics[width=0.16\linewidth]{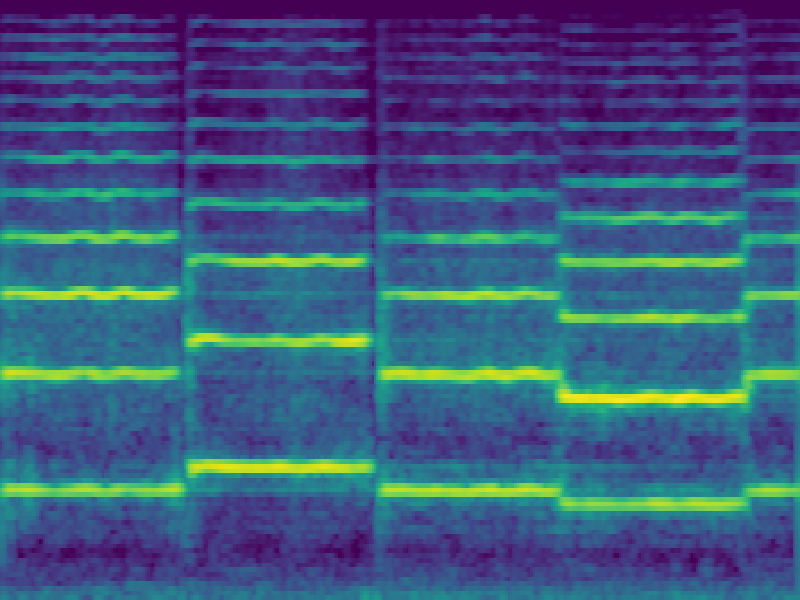} &
\includegraphics[width=0.16\linewidth]{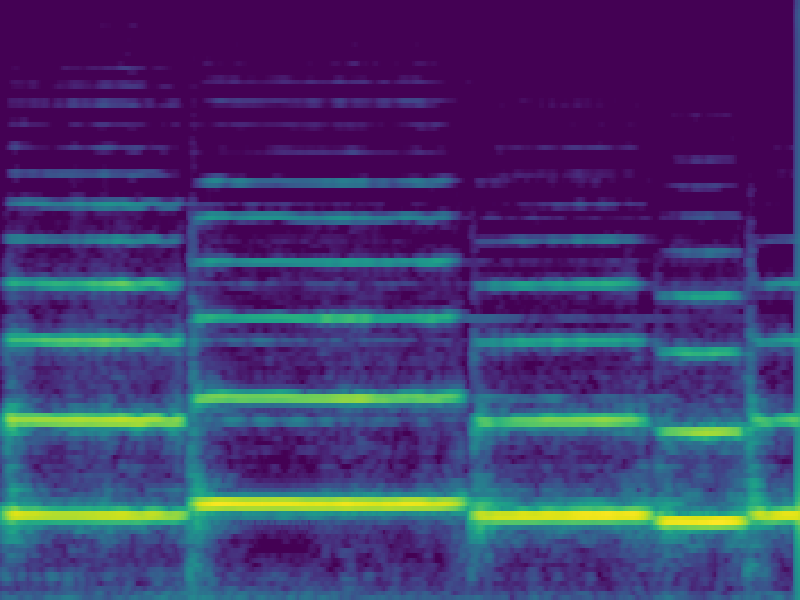} &
\includegraphics[width=0.16\linewidth]{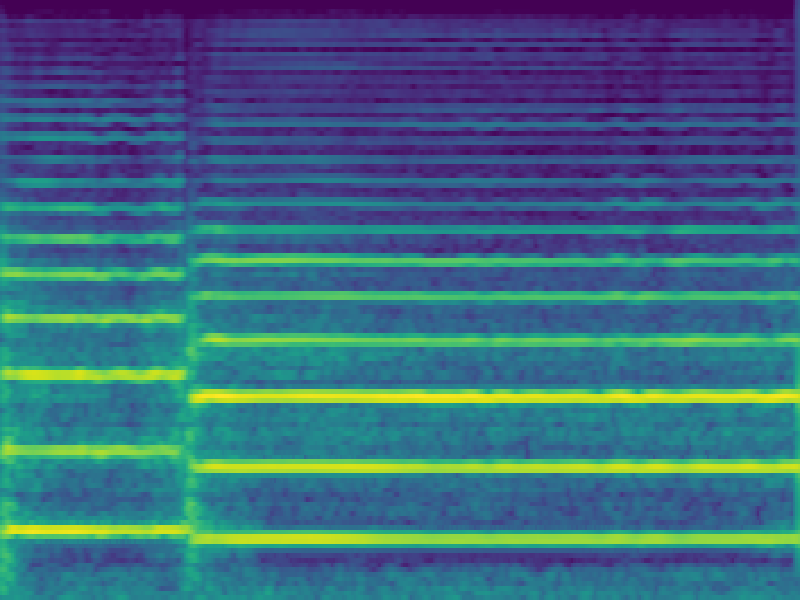} &
\includegraphics[width=0.16\linewidth]{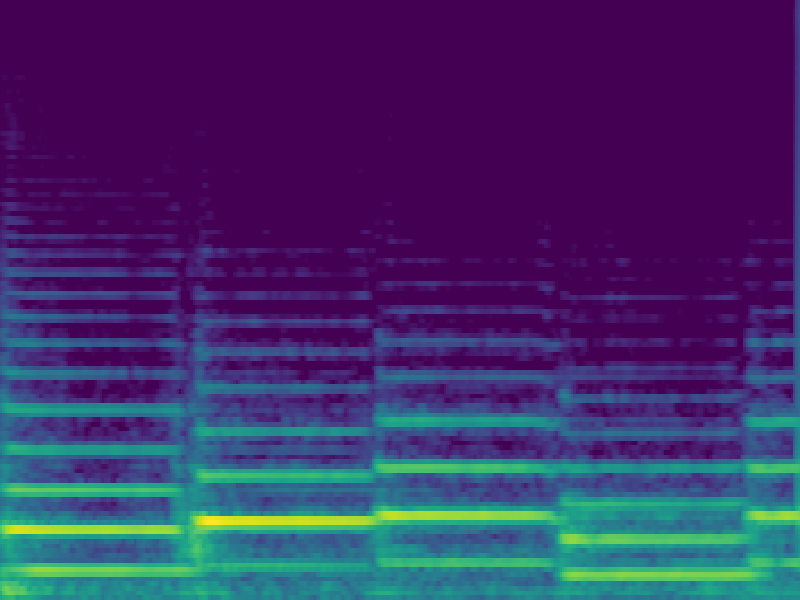} \\[2pt]
 & \scriptsize Gen.\ mix & \scriptsize Gen.\ trumpet & \scriptsize Gen.\ horn & \scriptsize Gen.\ trombone & \scriptsize Gen.\ tuba \\
\rotatebox{90}{\scriptsize\textbf{Output}} &
\includegraphics[width=0.16\linewidth]{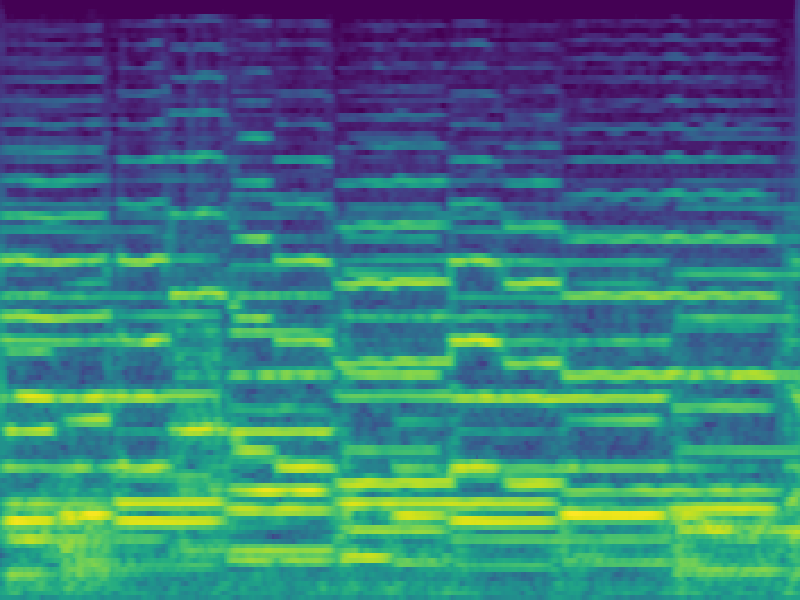} &
\includegraphics[width=0.16\linewidth]{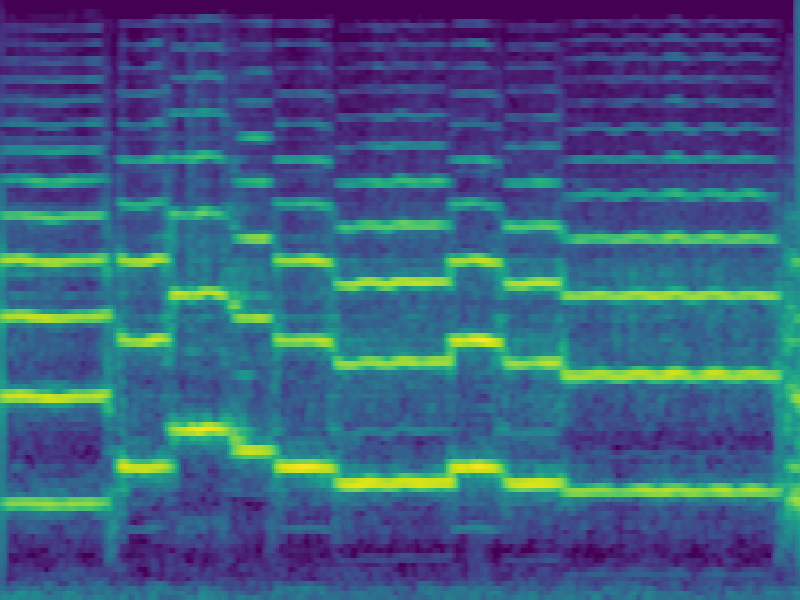} &
\includegraphics[width=0.16\linewidth]{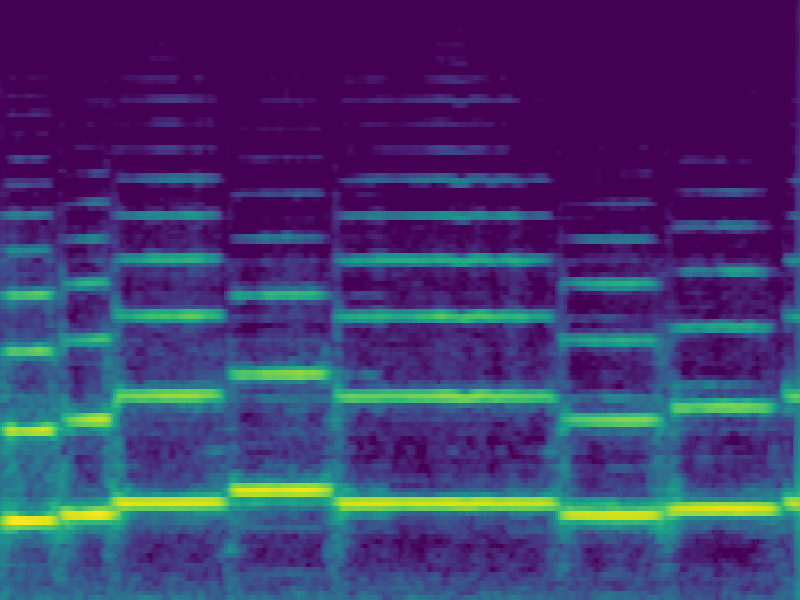} &
\includegraphics[width=0.16\linewidth]{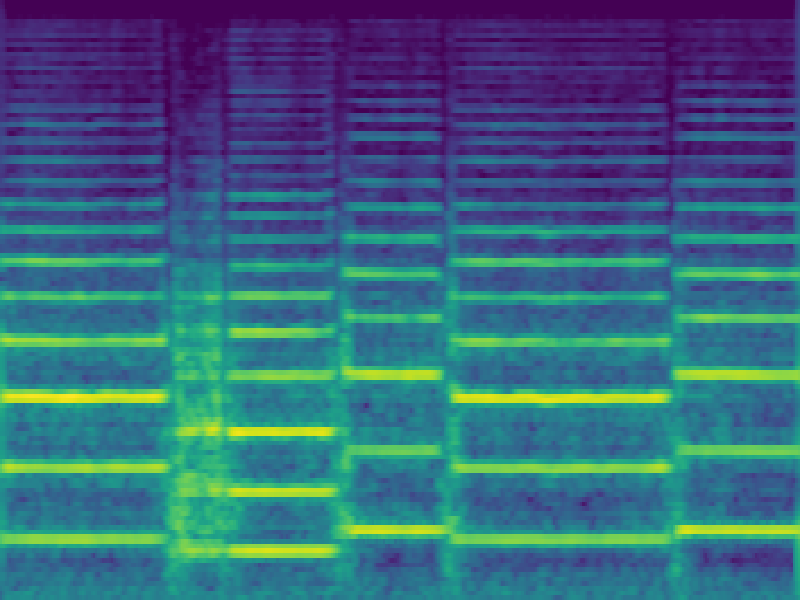} &
\includegraphics[width=0.16\linewidth]{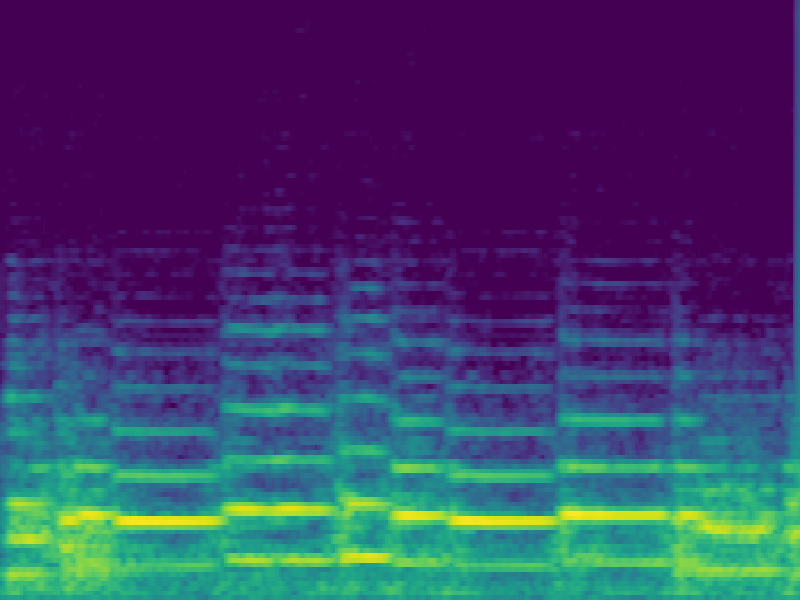} \\
\end{tabular}
\caption{Qualitative result of MixtureTT jointly generating from an input mixture and four timbre references.}
\label{fig:qualitative}
\end{figure}

Table~\ref{tab:main_results} shows that MixtureTT achieves the
best overall performance on CocoChorales, outperforming both
practical separate-then-transfer pipelines and the training-free
mixture baseline while remaining competitive with privileged methods
given isolated stems. These results demonstrate that direct mixture-to-stem timbre transfer can avoid materializing waveform-level source estimates without sacrificing source-level control.
The advantage persists on the real-audio ChoraleBricks benchmark,
where MixtureTT retains the strongest overall result profile. This
indicates that the proposed formulation generalizes beyond the
synthesized audio distribution used for training. A representative
transfer example is shown in Fig.~\ref{fig:qualitative}.

\subsection{Architectural and Mechanistic Analysis}
\label{ssec:ablation}

\paragraph{Interaction structure.}
\begin{table}[t]
\centering
\scriptsize
\setlength{\tabcolsep}{1.6pt}
\renewcommand{\arraystretch}{1.08}
\setlength{\aboverulesep}{1pt}
\setlength{\belowrulesep}{1pt}
\caption{
Architecture ablation on CocoChorales under reconstruction (R) and
cross-timbre transfer (T) settings.
}
\label{tab:architecture_ablation}
\resizebox{\columnwidth}{!}{%
\begin{tabular}{@{}l cc cc cc cc cc cc@{}}
\toprule
\multirow{3}{*}{\textbf{Architecture}} &
\multicolumn{8}{c}{\textbf{Per-Stem}} &
\multicolumn{4}{c}{\textbf{Mixture}} \\
\cmidrule(lr){2-9}\cmidrule(l){10-13}
&
\multicolumn{2}{c}{FAD$\downarrow$} &
\multicolumn{2}{c}{JD$\downarrow$} &
\multicolumn{2}{c}{MFCC$\downarrow$} &
\multicolumn{2}{c}{Conf$\uparrow$} &
\multicolumn{2}{c}{FAD$_m\!\downarrow$} &
\multicolumn{2}{c}{CCS$\uparrow$} \\
\cmidrule(lr){2-3}\cmidrule(lr){4-5}\cmidrule(lr){6-7}
\cmidrule(lr){8-9}\cmidrule(lr){10-11}\cmidrule(l){12-13}
& R & T & R & T & R & T & R & T & R & T & R & T \\
\midrule

Independent stems
& 0.011 & 0.029
& \underline{0.159} & \underline{0.245}
& 0.006 & \underline{0.037}
& 0.988 & 0.979
& 0.008 & \underline{0.019}
& \textbf{0.985} & \underline{0.933} \\

Full self-attention
& 0.011 & 0.029
& 0.166 & 0.287
& 0.006 & 0.039
& 0.988 & 0.970
& \underline{0.007} & 0.020
& \underline{0.985} & 0.920 \\

Mixture injection
& \underline{0.011} & \underline{0.028}
& 0.172 & 0.318
& \underline{0.006} & {0.039}
& \textbf{0.989} & \textbf{0.983}
& 0.008 & 0.021
& 0.983 & 0.892 \\

\cmidrule(lr){1-13}

\textbf{MixtureTT}
& \textbf{0.010} & \textbf{0.018}
& \textbf{0.157} & \textbf{0.199}
& \textbf{0.005} & \textbf{0.033}
& \underline{0.989} & \underline{0.979}
& \textbf{0.007} & \textbf{0.014}
& 0.985 & \textbf{0.937} \\

\bottomrule
\end{tabular}%
}
\end{table}
Table~\ref{tab:architecture_ablation} first tests the central
hypothesis that mixture-level timbre transfer benefits from jointly
modeling the generated stems. Relative to the otherwise matched
independent-stem variant, MixtureTT improves transfer accuracy,
content preservation, and mixture-level quality. Joint denoising is
therefore functionally important rather than merely a convenient way
to process multiple independent outputs.

However, allowing interaction alone is insufficient. Full
self-attention, which permits direct exchange among all noisy-stem
tokens, performs no better than independent generation and is
particularly weaker in content and harmonic preservation. This
suggests that unrestricted interaction introduces cross-talk instead
of capturing useful ensemble dependencies.

Mixture injection reveals a complementary failure mode. Its high
target-instrument confidence indicates that the codec mixture latent
contains strong timbral evidence, but its substantially worse JD and
CCS show that this entangled representation provides unreliable
source-specific structure. Together, these comparisons show that
cross-stem dependencies should be mediated through source-aware
content rather than through raw mixture features or direct exchange
among evolving stem states. MixtureTT implements this structured
coordination through the ECB, allowing all stems to share synchronized
ensemble context while preserving their private generation states.

\paragraph{Interaction order.}

\begin{table}[t]
\centering
\scriptsize
\setlength{\tabcolsep}{1.6pt}
\renewcommand{\arraystretch}{1.08}
\setlength{\aboverulesep}{1pt}
\setlength{\belowrulesep}{1pt}
\caption{
Ablation on the position of content-mediated coordination under reconstruction (R) and timbre transfer (T).
}
\label{tab:stage_order_ablation}
\resizebox{\columnwidth}{!}{%
\begin{tabular}{@{}l cc cc cc cc cc cc@{}}
\toprule
\multirow{3}{*}{\textbf{Stage Order}} &
\multicolumn{8}{c}{\textbf{Per-Stem}} &
\multicolumn{4}{c}{\textbf{Mixture}} \\
\cmidrule(lr){2-9}\cmidrule(l){10-13}
&
\multicolumn{2}{c}{FAD$\downarrow$} &
\multicolumn{2}{c}{JD$\downarrow$} &
\multicolumn{2}{c}{MFCC$\downarrow$} &
\multicolumn{2}{c}{Conf$\uparrow$} &
\multicolumn{2}{c}{FAD$_m\!\downarrow$} &
\multicolumn{2}{c}{CCS$\uparrow$} \\
\cmidrule(lr){2-3}\cmidrule(lr){4-5}\cmidrule(lr){6-7}
\cmidrule(lr){8-9}\cmidrule(lr){10-11}\cmidrule(l){12-13}
& R & T & R & T & R & T & R & T & R & T & R & T \\
\midrule

Cross-stem middle
& \textbf{0.010} & \underline{0.019}
& \underline{0.161} & \underline{0.237}
& \underline{0.006} & \underline{0.036}
& 0.987 & 0.969
& \underline{0.007} & \underline{0.015}
& \underline{0.984} & \underline{0.931} \\

Cross-stem last
& 0.011 & 0.021
& 0.173 & 0.286
& 0.006 & 0.039
& \underline{0.989} & \underline{0.977}
& 0.008 & 0.016
& 0.981 & 0.912 \\

\cmidrule(lr){1-13}

\textbf{Cross-stem first}
& \underline{0.010} & \textbf{0.018}
& \textbf{0.157} & \textbf{0.199}
& \textbf{0.005} & \textbf{0.033}
& \textbf{0.989} & \textbf{0.979}
& \textbf{0.007} & \textbf{0.014}
& \textbf{0.985} & \textbf{0.937} \\

\bottomrule
\end{tabular}%
}
\end{table}

Having established the need for structured cross-stem coordination,
we next examine when it should occur. Table~\ref{tab:stage_order_ablation}
shows a consistent deterioration as coordination is delayed. Moving
the shared interaction to the middle preserves much of the low-level
audio quality but weakens content and harmonic consistency, while
placing it last causes a larger degradation across both stem- and
mixture-level measures.

These results suggest that late coordination cannot reliably reconcile
representations that have already specialized independently. The
stems should instead establish a shared interpretation of the musical
event before devoting most model capacity to source-private
generation, supporting the proposed
\emph{coordinate-then-specialize} ordering.

\begin{figure}[t!]
    \centering
    \includegraphics[width=\columnwidth]{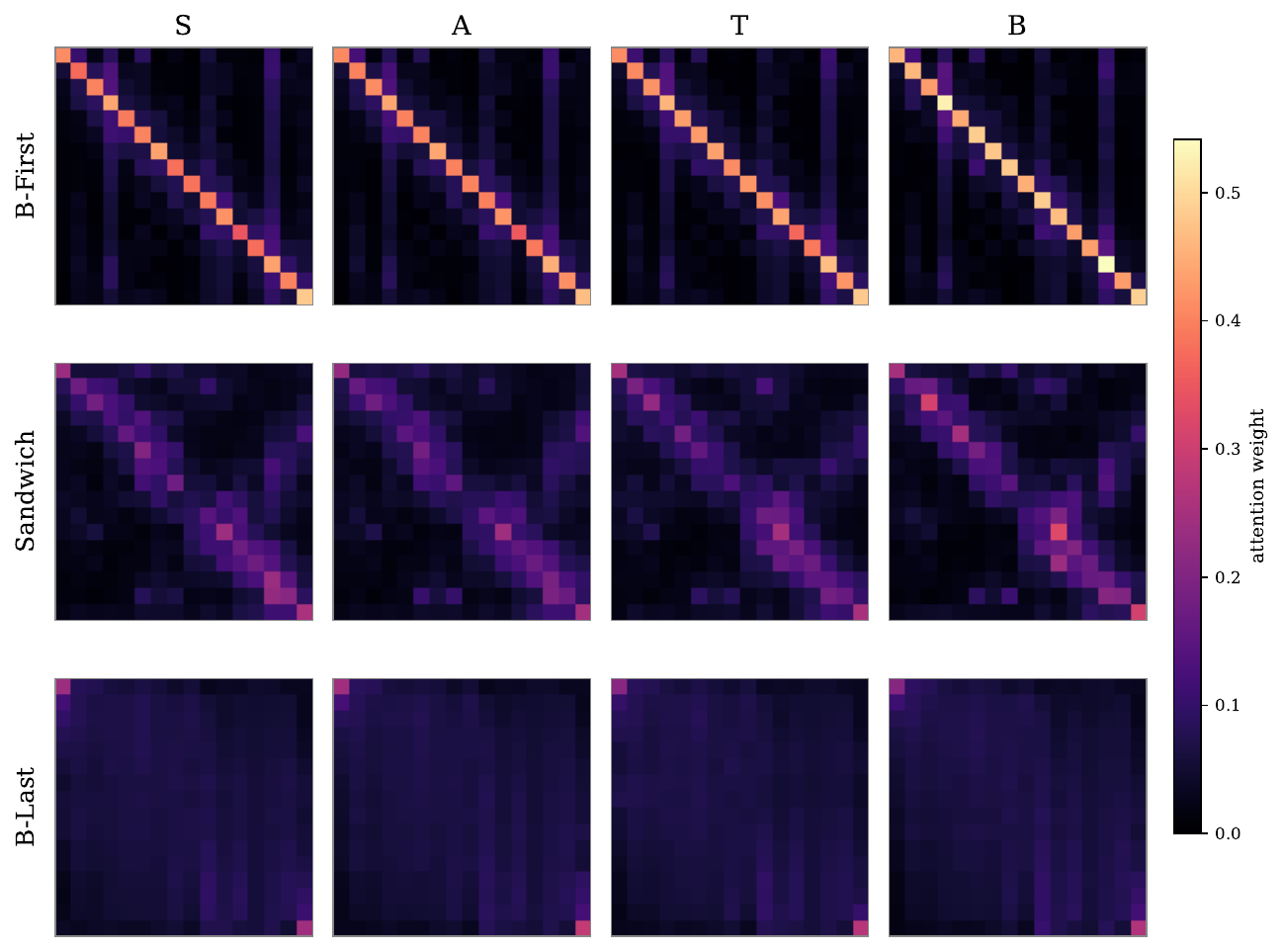}

    \caption{Cross-attention maps from noisy-stem queries to the ECB when coordination is placed early, middle, or late.
    }
    \label{fig:attention_visualization}
\end{figure}

Figure~\ref{fig:attention_visualization} provides a qualitative visualization consistent with this result. Rows index noisy-stem query times and
columns index ECB times. Because each ECB token aggregates all source
contents at the same time step, a diagonal pattern indicates that the
state at time $t$ retrieves synchronized ensemble context at time
$t$. This temporal alignment is clear under early coordination but
becomes increasingly diffuse when coordination is delayed, consistent
with private specialization making the shared structure harder to
recover.

\subsection{Disentanglement Analysis}
\label{ssec:disentanglement_results}

Table~\ref{tab:disentanglement_ablation} reveals that strong
structural scores do not necessarily indicate successful timbre
transfer. Without $\mathcal{L}_{\mathrm{cls}}$, JD and CCS remain
competitive, but transfer confidence drops sharply and audio quality
degrades. The model therefore preserves the musical structure while
largely retaining the original source identity, confirming that a
low-dimensional content bottleneck alone does not remove timbre
information.

Removing $\mathcal{L}_{\mathrm{div}}$ produces an even clearer
failure: transfer confidence approaches zero although several content
and mixture metrics improve. This apparent improvement is misleading,
as the collapsed target embeddings make the model insensitive to the
requested instruments and encourage input-like reconstruction.
Together, the two ablations show that both objectives are required for target-timbre control, rather than merely plausible audio
or content preservation.

\begin{table}[t]
\centering
\scriptsize
\setlength{\tabcolsep}{1.6pt}
\renewcommand{\arraystretch}{1.08}
\setlength{\aboverulesep}{1pt}
\setlength{\belowrulesep}{1pt}
\caption{
Ablation of the disentanglement objectives under reconstruction (R) and cross-timbre transfer (T).
}
\label{tab:disentanglement_ablation}
\resizebox{\columnwidth}{!}{%
\begin{tabular}{@{}l cc cc cc cc cc cc@{}}
\toprule
\multirow{3}{*}{\textbf{Objective}} &
\multicolumn{8}{c}{\textbf{Per-Stem}} &
\multicolumn{4}{c}{\textbf{Mixture}} \\
\cmidrule(lr){2-9}\cmidrule(l){10-13}
&
\multicolumn{2}{c}{FAD$\downarrow$} &
\multicolumn{2}{c}{JD$\downarrow$} &
\multicolumn{2}{c}{MFCC$\downarrow$} &
\multicolumn{2}{c}{Conf$\uparrow$} &
\multicolumn{2}{c}{FAD$_m\!\downarrow$} &
\multicolumn{2}{c}{CCS$\uparrow$} \\
\cmidrule(lr){2-3}\cmidrule(lr){4-5}\cmidrule(lr){6-7}
\cmidrule(lr){8-9}\cmidrule(lr){10-11}\cmidrule(l){12-13}
& R & T & R & T & R & T & R & T & R & T & R & T \\
\midrule

w/o $\mathcal{L}_{\mathrm{cls}}$
& 0.091 & 0.272
& \underline{0.141} & \underline{0.153}
& 0.012 & \underline{0.051}
& 0.988 & \underline{0.356}
& 0.063 & 0.294
& \textbf{0.991} & \underline{0.980} \\

w/o $\mathcal{L}_{\mathrm{div}}$
& \underline{0.088} & \underline{0.087}
& \textbf{0.138} & \textbf{0.139}
& \underline{0.011} & 0.073
& \textbf{0.991} & 0.001
& \underline{0.058} & \underline{0.070}
& \underline{0.990} & \textbf{0.990} \\

\cmidrule(lr){1-13}

\textbf{MixtureTT}
& \textbf{0.010} & \textbf{0.018}
& 0.157 & 0.199
& \textbf{0.005} & \textbf{0.033}
& \underline{0.989} & \textbf{0.979}
& \textbf{0.007} & \textbf{0.014}
& 0.985 & 0.937 \\

\bottomrule
\end{tabular}%
}
\end{table}

\subsection{Subjective Evaluation}
\label{ssec:subjective_results}

\begin{figure}[t!]
    \centering
    \includegraphics[width=\columnwidth]{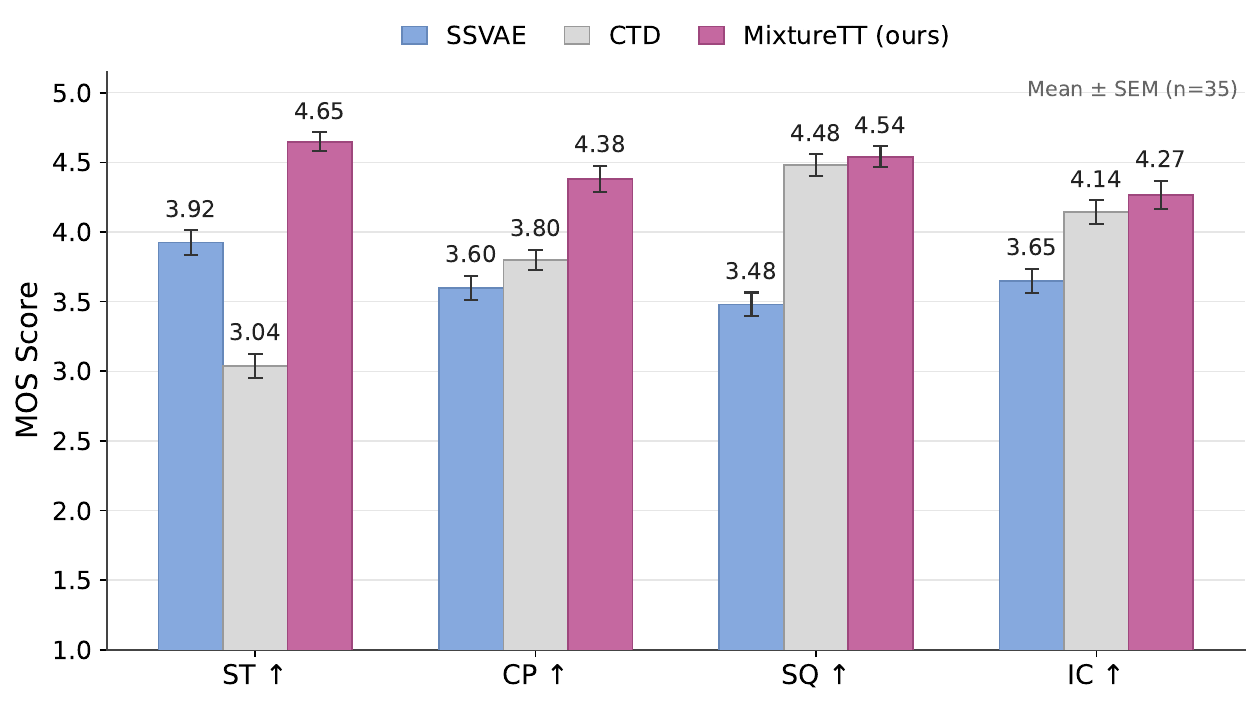}
    \caption{
Subjective Mean Opinion Scores (MOS; 1--5) for transfer success
(ST), content preservation (CP), sound quality (SQ), and inter-stem
coherence (IC).
}
    \label{fig:mos_results}
\end{figure}

Figure~\ref{fig:mos_results} shows that MixtureTT receives the highest
perceptual rating on all four criteria, despite the competing systems
being given ground-truth isolated stems. The gain in inter-stem
coherence is particularly important: listeners prefer the jointly
generated ensemble to the mixdown of independently transferred stems,
providing direct perceptual evidence that modeling cross-stem
dependencies benefits mixture-level timbre transfer. The concurrent
improvements in transfer success and sound quality further show that
this coherence is not obtained by weakening per-stem timbre control.

\subsection{Scaling with Pseudo-Labeled Data}
\label{ssec:scaling}
Table~\ref{tab:data_scaling} reports performance as paired
CocoChorales examples are progressively replaced by mixtures
pseudo-labeled with HT Demucs. Audio-quality and timbre metrics remain
relatively stable as paired supervision decreases, while content and
harmonic preservation degrade gradually. Even training without paired
ground-truth stems produces usable outputs, suggesting that MixtureTT
can reduce its dependence on manually paired multitrack data.

\begin{table}[t!]
\centering
\caption{Effect of supervision and data scale. $\mathcal{D}_s$: paired data; $\mathcal{D}_u$: external unpaired data.}
\label{tab:data_scaling}
\renewcommand{\arraystretch}{1.1}
\resizebox{\columnwidth}{!}{%
\begin{tabular}{@{}cc|cc|cccc|cc@{}}
\toprule
\multicolumn{2}{c|}{\textbf{Sup.}}
  & \multicolumn{2}{c|}{\textbf{Unsup.}}
  & \multicolumn{4}{c|}{\textbf{Stem}}
  & \multicolumn{2}{c}{\textbf{Mix}} \\
\cmidrule(lr){5-8}\cmidrule(l){9-10}
& & & & FAD$\downarrow$ & JD$\downarrow$ & MFCC$\downarrow$ & Conf$\uparrow$ & FAD$_m\!\downarrow$ & CCS$\uparrow$ \\
\midrule
$\mathcal{D}_s$ & 100\% & --- & 0\%
  & 0.018 & 0.199 & 0.033 & 0.979 & 0.014 & 0.937 \\
$\mathcal{D}_s$ & 50\% & $\mathcal{D}_u$ & 50\%
 & 0.018 & 0.231 & 0.031 & 0.972 & 0.015 & 0.909 \\
$\mathcal{D}_s$ & 10\% & $\mathcal{D}_u$ & 90\%
  & 0.019 & 0.264 & 0.032 & 0.964 & 0.016 & 0.882 \\
$\mathcal{D}_s$ & 5\% & $\mathcal{D}_u$ & 95\%
  & 0.020 & 0.271 & 0.033 & 0.958 & 0.016 & 0.871 \\
--- & 0\% & $\mathcal{D}_u$ & 100\%
  & 0.027 & 0.241 & 0.034 & 0.945 & 0.016 & 0.909 \\
\bottomrule
\end{tabular}%
}
\end{table}

\section{Conclusion}
\label{sec:conclusion}
This paper presented MixtureTT, to our knowledge the first joint latent diffusion system to flexibly re-instrument individual voices directly within a polyphonic mixture, without explicit waveform-level source separation, source-identification queries, or instrument labels. MixtureTT generates all target stems within a shared diffusion process, explicitly modeling their dependencies while preserving source-specific content and independent target-timbre control. Despite operating under a strictly harder input condition, the proposed Joint Stem Diffusion Transformer outperforms strong single-instrument baselines, while controlled ablations confirm cross-stem modeling as a key factor underlying this advantage. 

Current limitations include the fixed four-source setting and dependence on a pretrained source-aware backbone. Future work will extend MixtureTT to variable-size ensembles, scale training with larger collections of pseudo-labeled mixtures, and further investigate the relationship between content--timbre disentanglement and ensemble coherence. More broadly, we hope this work 
initiates a reflection on treating the polyphonic mixture as 
the natural object of study in generative music modeling, 
encouraging further exploration of mixture-level formulations 
across the broader family of music-manipulation tasks.

\bibliography{references}

\end{document}